# Fabrication of thin diamond membranes for photonic applications

Jonathan C. Lee, Andrew P. Magyar, David O. Bracher, Igor Aharonovich* and Evelyn L. Hu

School of Engineering and Applied Sciences, Harvard University, Cambridge. 02138, MA, USA

* igor@seas.harvard.edu

**Abstract**
High quality, thin diamond membranes containing nitrogen-vacancy centers provide critical advantages in the fabrication of diamond-based structures for a variety of applications, including wide field magnetometry, photonics and bio-sensing. In this work we describe, in detail, the generation of thin, optically-active diamond membranes by means of ion implantation and overgrowth. To establish the suitability of our method for photonic applications, photonic crystal cavities with quality factor of 1000 are fabricated.

## 1. Introduction

Nitrogen-vacancy (NV) centers in diamond have emerged as one of the most promising solid state qubits due to their room temperature operation, long spin coherence times, and suitability for optical initialization and readout[1-6]. Beyond their quantum photonic applications, NV centers in diamond have been recently utilized to demonstrate electric and magnetic field sensing with exceptional sensitivity and spatial resolution [7-10]. Furthermore, the bio-compatibility and the chemical inertness of diamond enable applications of NVs in diamond as biosensors for cellular and neural activities[11-13].

Many applications in the field of photonic technologies and bio-sensing require availability of high quality, thin diamond membranes with stable NV centers. For instance, coupling single emitters to photonic crystal cavities (PCCs) or waveguides requires thin (~ 300 nm) diamond membranes having only a few NV centers.[14, 15] Alternatively, recent schemes for ensemble magnetometry demand high concentrations of NV centers in a free-standing, single crystal diamond slab[10, 16]. Thin diamond membranes with diluted NV centers provide an excellent platform for the formation of optical cavities, strongly coupled to single emitters.

Several approaches have been utilized to form free-standing diamond membranes from bulk single-crystal diamond. One approach involves high energy (~ a few MeV) and high dose ion implantation (~ $1\times10^{17}$ ions/cm$^2$) to selectively damage a portion of the diamond, sub-surface, followed by a lift-off of the membrane [17]. Other approaches employ focused ion beam (FIB) milling to etch or 'cut out' suspended diamond membranes[18, 19] and vertical etch of diamond substrates[20]. Unfortunately, in both cases, the process of creating the membrane introduces ion damage that degrades the optical performance of the NVs within the membrane. In addition, a residual built-in strain, resulting from the high ion damage, makes subsequent processing of the membranes extremely challenging [21]. Recently, our group demonstrated that an overgrowth method dramatically improves the optical properties of the membranes, making them suitable for device engineering [22, 23]. Subsequent reports have shown that a short overgrowth of diamond on a bulk diamond template can improve coherence times of implanted NVs as well improve the stability of near-surface NV centers [24].

In this paper we detail a methodology to generate high quality thin diamond membranes that are suitable for photonic applications. To demonstrate the quality and utility of this material, we fabricate a PCC with quality factor of ~ 1000 from the diamond membrane.

## 2. Results

Figure 1 shows a schematic of our process. A type IIA CVD single crystal diamond is subject to ion implantation (1MeV, $5 \times 10^{16}$ He/cm$^2$) and subsequently annealed for one hour at 900°C under nitrogen flow. A hard mask for reactive ion etching (RIE) is created by depositing 300 nm of $SiO_2$ on the sample via plasma-enhanced chemical vapour deposition (PECVD). Square mesas (~250 × 250 μm$^2$) are patterned using optical lithography, and the pattern is transferred to the $SiO_2$ and subsequently to the diamond, utilizing a fluorine based RIE followed by an $O_2$-RIE. Finally, the sample is immersed in hydrofluoric acid to remove the oxide layer.

To generate membranes, the diamond is immersed in ultrapure water (Millipore) and the damaged region is etched electrochemically under constant potential conditions, resulting in floating diamond membranes. Using a pipette tip, the membranes are picked up from the solution and placed on a substrate of choice. The membranes have (100) crystallographic orientation, the same as the original type IIA diamond crystal.

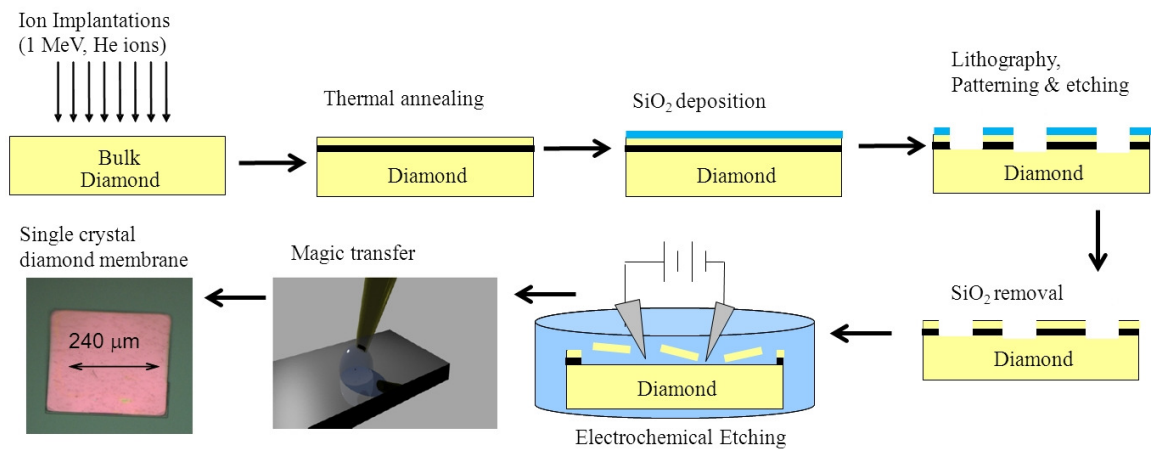

*Figure 1. Schematic illustration of the procedure used to generate diamond membranes. First, a CVD diamond crystal is implanted with He$^+$ ions to generate a damaged region, 1.7 μm beneath the surface of the diamond. The sample is annealed and patterned to lithographically define the membrane size. Reactive ion etching is employed to transfer the pattern from the mask to the diamond and to subsequently etch the diamond. Electrochemical etching enables the lift-off of the membranes and a pipette is used to transfer the membrane to a substrate of choice.*

The original implantation depth, which determines the thickness of the initially-formed membranes, is chosen to be 1.7 μm so that the most heavily damaged material is ultimately removed during processing [21]. The residual damage in the membrane produces a built-in strain, which results in curving of the membranes. The large curvature of the diamond membranes in turn limits further processing, including the thinning of the membrane and the uniform removal of the damaged material. To minimize membrane curvature, the membrane is mechanically stamped onto a poly-methyl methacrylate (PMMA) spin coated substrate. The PMMA acts as a glue layer and holds the membrane flat. The process is shown schematically in Fig 2, with the curved membrane after the electrochemical etch and after being stamped on the PMMA and thinned using $O_2$-RIE. Subsequent thinning of the membrane reveals a membrane with a thickness of ~ 190 nm.

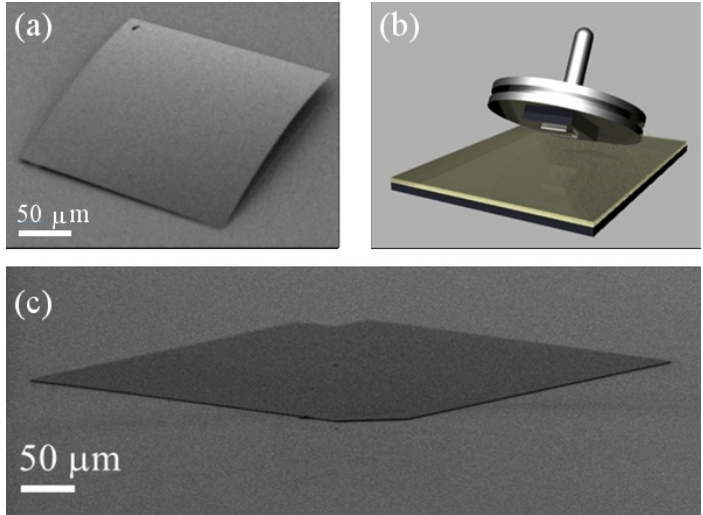

*Figure 2. (a) Diamond membrane exhibiting curvature. (b) A schematic of the stamping process. The diamond membrane is mounted on a sample holder and stamped onto a substrate coated with PMMA. (c) A stamped and thinned membrane exhibiting uniform thickness without noticeable curvature.*

Once the membrane is transferred onto $SiO_2$ substrate, optical cavities (microdisks and microrings) are patterned by using a 125 kV electron beam writer (Elionix ELS-F125) with 50 nm SiO2 hard mask on top of the diamond membrane. Fig 3a,b shows SEM images of the microdisk and the microring, respectively, fabricated from the original diamond membranes. The optical resonators are fabricated by etching the diamond membranes using $O_2$-RIE. Figs. 3c, d show the corresponding PL spectra recorded from microdisk and microring cavities using a 532 nm excitation laser at room temperature. The spectra are collected under a confocal microcope using a 100x objective with a 0.95 numerical aperture. The peaks in the emission spectrum are whispering gallery mode (WGMs) resonances from the optical cavities. The WGMs exhibit quality factors, $Q$, of approximately 500 and 1000 for the microdisk and the microring, respectively. We believe that the magnitude of $Q$ is reduced by losses to defects in the membrane material, resulting from residual damage from the ion implantation damage. Such residual damage is evident from the Raman measurements[21]. The $Q$ of the microdisk is comparable with that of optical cavities fabricated by FIB[18].

To further improve the quality of the optical resonators, a diamond overgrowth step is introduced into the fabrication process. After the original membrane is generated, a thin layer of diamond is epitaxially regrown using a conventional microwave CVD reactor. The overgrowth conditions are 1% $CH_4$ in 400 SCCM of $H_2$ at a pressure of 60 torr for approximately 10 minutes. We note that because no heating stage is used during the growth the temperature is determined by the plasma density. After the overgrowth, the composite membrane is flipped and the damaged membrane template is fully etched away, leaving behind a pristine, CVD grown diamond slab – only a few hundred nanometers thick.

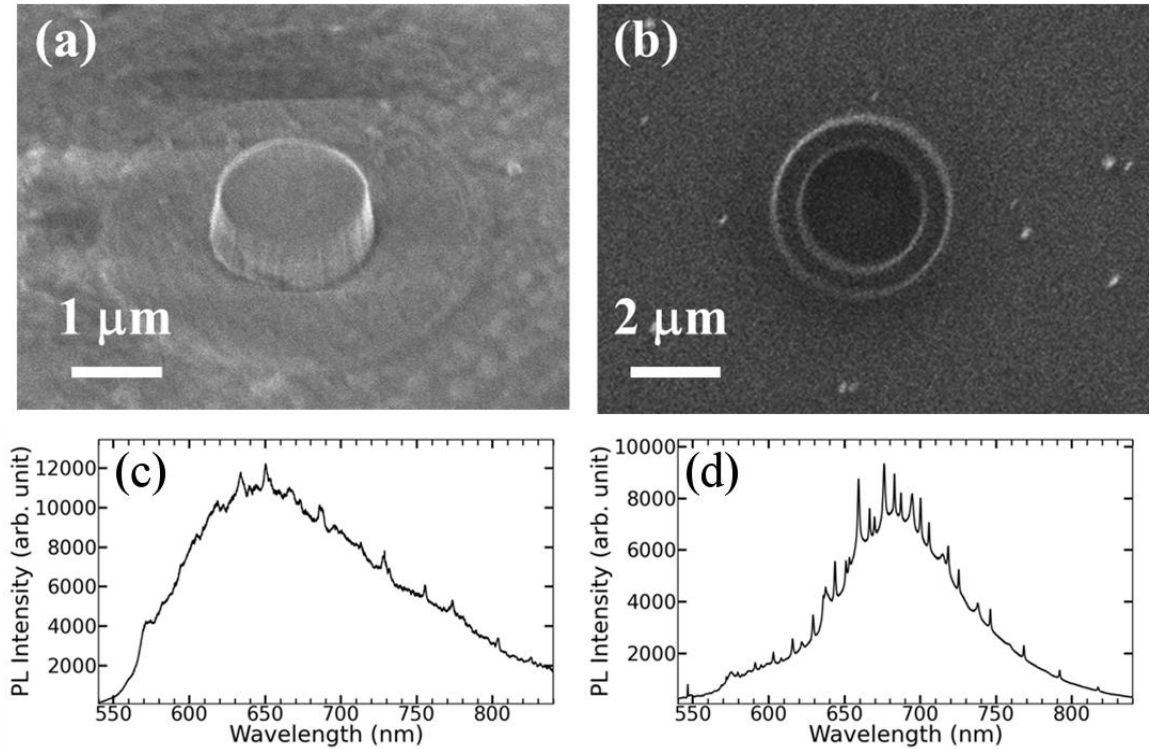

*Figure 3. SEM image of a diamond microdisk (a) and diamond microring (b) fabricated from the original diamond membrane. (c) PL spectrum from a 1.6 μm diameter microdisk with whispering gallery modes (Q ~ 500) (d) PL spectrum from a 3.6 μm diameter, 1.2 μm width microring resonator shows whispering gallery modes with Q ~ 1000.*

Photonic crystal cavities are fabricated from the thin diamond membrane formed by overgrowth. Figure 4a shows a SEM image of the PCC fabricated in the diamond membrane. Figure 4b shows a magnified image of the central region of the H1 PCC, having hexagonal symmetry with a single hole photonic crystal 'defect'[25]. Fig 4b shows a PL spectrum recorded from the PCC using a 532 nm excitation source at room temperature. The broad PL emission is decorated by narrow lines, which are the modes of the PCC. Fig 4c shows a high-resolution spectrum of one of the modes, with $Q$ ~ 1000. Although the $Q$ is still lower than state of the art diamond resonators[14, 15, 26], it is a significant improvement over devices fabricated from ion implanted and non-regrown membranes. Losses in these PCC devices can be due to imperfections in fabrication, primarily during the diamond-etching step. Additionally, the underlying PMMA layer is not fully undercut and therefore limits vertical confinement. Finally, we note that microdisk cavities fabricated from the regrown membranes exhibit high quality factors of up to 3000 [22, 23].

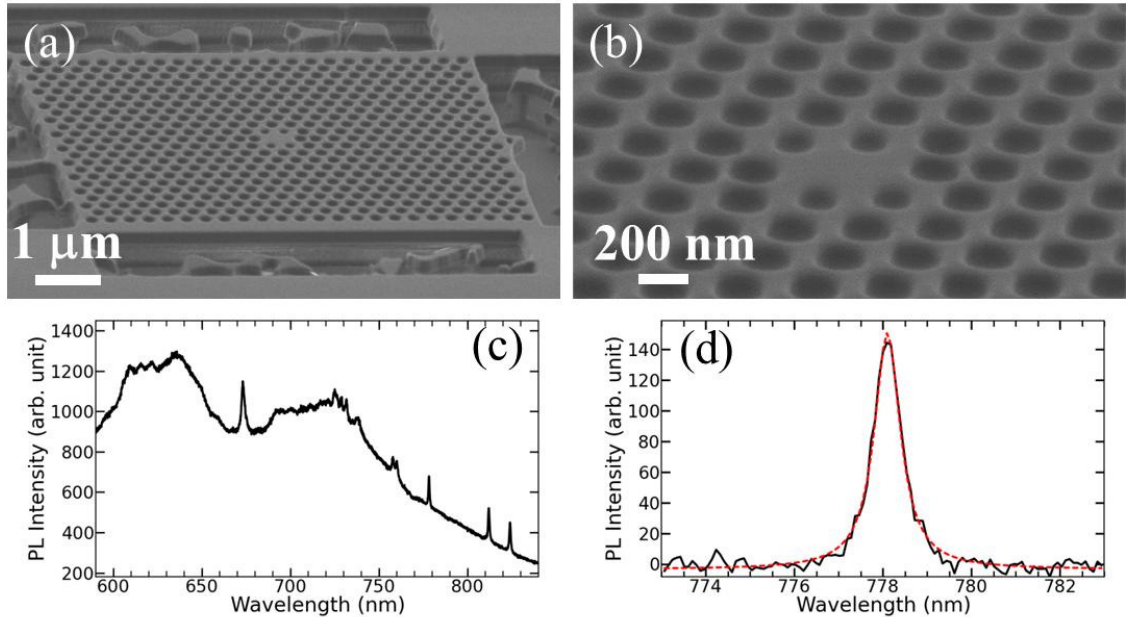

*Figure 4. (a) SEM image of the H1 PCC with 65 degree tilted angle. b.) zoomed in image of the H1 PCC with 65 degree tilted angle. c.) PL spectra from the cavity show modes. d.) zoomed in PL spectra of the cavity mode with Q ~ 1000. The black line is the PL curve and the dotted red line is the fitted curve.*

## 3. Conclusions

To summarize, we have presented in detail the generation of high quality, thin diamond membranes that are suitable for a variety of applications in photonics, sensing and magnetometry. Ion implantation is employed to generate diamond templates and overgrowth is utilized to achieve high quality diamond membranes. Demonstrating their utility, a PCC with a quality factor of 1000 is fabricated from these membranes. These diamond membranes can find applications in wide field imaging, sensing and ensemble magnetometry as they contain a large number of NV centers. Alternatively, controlling the nitrogen incorporation by using appropriate growth conditions can result in isotopically pure diamond [24, 27, 28] with single emitters – advantageous for quantum photonic applications. The developed overgrowth process is a powerful method to generate high quality optically and spin active thin diamond membranes.

**Acknowledgments**
The authors acknowledge the help of D.R. Clarke for access to the PL and Raman facilities, and M. Huang for assistance with ion implantation. This work was carried out with the financial support of DARPA under the Quantum Entanglement Science and Technology (QuEST) Program. This work was enabled by facilities available at the Center for Nanoscale Systems (CNS), a member of the National Nanotechnology Infrastructure Network (NNIN), which is supported by the National Science Foundation under NSF award no. ECS-0335765. CNS is part of the Faculty of Arts and Sciences at Harvard University.